\documentclass[journal]{IEEEtran}
\usepackage[utf8]{inputenc}
\usepackage{color}
\usepackage{xcolor}
\usepackage{array}
\usepackage{verbatim}
\usepackage{float}
\usepackage{amsmath}
\usepackage{amsthm}
\usepackage{amssymb}
\usepackage{graphicx}
\usepackage{longtable}
\usepackage{multirow}
\usepackage{booktabs}
\usepackage[unicode=true,
bookmarks=false,
breaklinks=false,pdfborder={0 0 1},colorlinks=false]
{hyperref}
\hypersetup{
	colorlinks,bookmarksopen,bookmarksnumbered,citecolor=blue,urlcolor=blue}
\usepackage{cite}

\usepackage{array}
\usepackage{overpic}
\usepackage{dblfloatfix}% To enable figures at the bottom of page
\usepackage{balance}

\usepackage{lipsum}
\usepackage{mathtools}
\usepackage{cuted}
\providecommand{\tabularnewline}{\\}
\usepackage{algorithmic}
\usepackage{longtable}

\newcommand{\Pexp}{\mathbin{\text{$\vcenter{\hbox{\textcircled{$e$}}}$}}}

\floatstyle{ruled}
\newfloat{algorithm}{tbp}{loa}
\providecommand{\algorithmname}{Algorithm}
\floatname{algorithm}{\protect\algorithmname}

\makeatletter
\let\oldforeign@language\foreign@language
\DeclareRobustCommand{\foreign@language}[1]{%
	\lowercase{\oldforeign@language{#1}}}

\let\oldforeign@language\foreign@language
\DeclareRobustCommand{\foreign@language}[1]{%
	\lowercase{\oldforeign@language{#1}}}

\ifCLASSINFOpdf
\else
\fi

\@ifundefined{showcaptionsetup}{}{%
	\PassOptionsToPackage{caption=false}{subfig}}
\usepackage{subfig}

\usepackage{balance}

\ifCLASSINFOpdf
\else
\fi

\newtheorem{thm}{Theorem}
\newtheorem{rem}{Remark}

\newtheorem{assum}{Assumption}
\ifCLASSINFOpdf
\else
\fi

\def\ps@IEEEtitlepagestyle{%
	\def\@oddhead{\parbox[t][\height][t]{\textwidth}{\centering \scriptsize
			Personal use of this material is permitted. Permission from the author(s) and/or copyright holder(s), must be obtained for all other uses. Please contact us and provide details if you believe this document breaches copyrights.\\
			\noindent\makebox[\linewidth]{}
		}\hfil\hbox{}}%
	\def\@evenhead{\scriptsize\thepage \hfil \leftmark\mbox{}}%
	\def\@oddfoot{\parbox[t][\height][l]{\textwidth}{
			\vspace{-20pt}{\rule{\textwidth}{0.4pt}}\\ \footnotesize{\bf{\footnotesize\textcolor{red}{H. Naser, H. A. Hashim, and M. Ahmadi, "Human Physical Interaction based on UAV Cooperative Payload Transportation System using Adaptive Backstepping and FNTSMC," In Proc. of the 2025 IEEE American Control Conference, Denver, Colorado, USA 2025.}}}\\
			\noindent\makebox[\linewidth]
		}\hfil\hbox{}}%
	\def\@evenfoot{\MYfooter}}

\makeatother
\begin{document}
	\bstctlcite{IEEEexample:BSTcontrol}

\title{Human Physical Interaction based on UAV Cooperative Payload Transportation System using Adaptive Backstepping and FNTSMC}

\author{Hussein Naser, Hashim A. Hashim, and Mojtaba Ahmadi% <-this % stops a space
	\thanks{This work was supported in part by the National Sciences and Engineering Research Council of Canada (NSERC), under the grants RGPIN-2022-04937.}
	\thanks{H. Naser, H. A. Hashim, and M. Ahmadi are with the Department of Mechanical
		and Aerospace Engineering, Carleton University, Ottawa, Ontario, K1S-5B6,
		Canada (e-mail: hhashim@carleton.ca).}
}

% \markboth{IEEE TRANSACTIONS ON INTELLIGENT TRANSPORTATION SYSTEMS, \today}{Hashim \MakeLowercase{\textit{et al.}}: Landmark and IMU Data Fusion: Systematic Convergence Geometric Nonlinear Observer for SLAM and Velocity Bias}

% \markboth{}{Hashim \MakeLowercase{\textit{et al.}}: Nonlinear Filter for Simultaneous Localization and Mapping on a Matrix Lie Group using IMU and Feature Measurements}

\maketitle
\begin{abstract}
This paper presents a nonlinear control strategy for an aerial cooperative
payload transportation system consisting of two quadrotor UAVs rigidly
connected to a payload. The system includes human physical interaction
facilitated by an admittance control. The proposed control framework
integrates an adaptive Backstepping controller for the position subsystem
and a Fast Nonsingular Terminal Sliding Mode Control (FNTSMC) for
the attitude subsystem to ensure asymptotic stabilization. The admittance
controller interprets the interaction forces from the human operator,
generating reference trajectories for the position controller to ensure
accurate tracking of the operator’s guidance. The system aims to assist
humans in payload transportation, providing both stability and responsiveness.
The robustness and effectiveness of the proposed control scheme in
maintaining system stability and performance under various conditions
are presented.
\end{abstract}

\section{Introduction}\label{sec1}

\IEEEPARstart{A}{erial} cooperative payload transportation using multiple UAVs is a
growing field with significant potential in various applications \cite{hash2025_RIENG_Avionics},
such as disaster relief scenarios, construction and industrial operations,
and agriculture \cite{real2021experimental,hashim2023exponentially,hachiya2022reinforcement,hash2025_RIENG_Avionics}.
Robust adaptive estimators and control techniques for UAVs addressing
uncertainties, ensuring synchronization, and rejecting disturbances
are in high demand \cite{hash2025_RIENG_Avionics,hash2024_TITS_UWB}.
Researchers have investigated different methods for grasping and transporting
payloads cooperatively, such as suspended payloads \cite{9540662}
and aerial grasping manipulators \cite{10178351,lee2018integrated}.
For suspended payloads, researchers in \cite{9540662}, developed
a two quadrotors system to transport a beam-shaped payload. The quadrotors
and the payload were considered as a single system and modeled using
virtual structure approach. On the other hand, in \cite{10178351}
a leader-follower setting was implemented for cooperative transportation
of payloads by a team of aerial manipulators. In the leader-follower
paradigm, the leader is preprogrammed to navigate through a predefined
path, while all other followers attempt to maintain their relative
pose with respect to the leader. In \cite{lee2018integrated}, a cooperative
transportation system using multiple aerial manipulators was developed.
An adaptive controller and an online estimation method were proposed
to estimate the mass of an unknown payload and control the aerial
manipulation system based on estimated parameters.

The ability to control multiple UAVs cooperatively to transport and
manipulate payloads presents numerous challenges, particularly in
ensuring precise control, stability, and safety \cite{tan2018cooperative,ref19,hashim2023exponentially,naser2025aerial,hash2024_TNN_Att_ObsvCont}.
These challenges become even more complex when considering the physical
interactions between the aerial vehicle and human operators. In \cite{romano2022cooperative},
the authors proposed a haptic slung payload system that consists of
five quadrotors connected to the payload by cables. The system can
move the payload according to human user guidance in 3D space by applying
forces to it. The system relies on inline tension sensors for force
estimation using PID and admittance control to track the user's haptic
guidance. In \cite{li2022safety,li2021cooperative}, a cooperative
suspended-payload transportation system was developed. The system
consists of a rigid triangle-shaped payload transported by multiple
micro aerial vehicles (MAVs) using massless cables. The authors implemented
an admittance controller for the human-robot interaction and a hierarchical
nonlinear controller to control the multiple MAVs during the transportation
and manipulation tasks. \vspace*{0.2cm}

To control cooperative payload transportation missions, researchers
used PID controllers \cite{barawkar2019cooperative}. Linear control
methods often fail due to the system nonlinear nature, the presence
of external disturbances, and the need for real-time adaptability
to change in environment \cite{hash2025_RIENG_Avionics,hashim2023exponentially,hashim2023observer,hash2021_AeScTe_SLAM}.
Also, incorporating human-physical interaction adds another layer
of complexity to the control system. As a remedy, robust nonlinear
control such as sliding mode control, backstepping, and adaptive control
were investigated to stabilize the system, compensate for system uncertainties,
reject external disturbances, and accommodate human physical interaction
\cite{xiong2017global,labbadi2020robust,naser2025aerial}. Although
using cables in cooperative payload transportation is popular, low-cost,
and simple to implement, it suffers from some limitations and drawbacks.
Cable-suspended payloads oscillate during the transportation task,
which affects the stability of the system and necessitates complex
control algorithms to stabilize the system \cite{naser2025aerial}.
In addition, oscillation interferes with the human interaction input
and makes it difficult to implement the admittance controller. Furthermore,
cables and tethers can get entangled during the task and destabilize
the system, causing potential system damage and decreasing the safety
level of the human partner. On the other hand, using aerial manipulators
has the advantage of grasping and manipulating the payload autonomously;
however, it suffers from serious drawbacks such as high cost, energy
consumption, and the added mass of the robotic arms.

\vspace*{0.2cm}

\paragraph*{Contribution}

Motivated by the above challenges, in this work a rigidly connected
cooperative transportation payload system consisting of two quadrotors
and a transported payload with human physical interaction is proposed.
The rigid connection is advantageous compared to other configurations;
It unifies the quadrotors dynamics and the payload as one rigid-system,
eliminates the oscillation to provide precise measurements of the
physical interactions, and increases the operator safety. For the
system's control, the proposed approach combines the strengths of
Backstepping control to ensure position subsystem stability, Fast
Nonsingular Terminal Sliding Mode Control (FNTSMC) for robust performance
and rapid attitude subsystem convergence, and the admittance controller
to facilitate the intuitive human physical interaction. The integration
of these control strategies enhance the overall system performance
and stability while addressing the cooperative payload transportation
key challenges. \vspace*{0.3cm}

The paper is organized as follows: Section \ref{methodology} details
the methodology, including preliminaries, problem formulation, and
system dynamics. Section \ref{control_section} presents the position,
attitude, and admittance control designs and stability analysis. The
simulation and numerical results are presented in Section \ref{results-discussion}.
Finally, the conclusion is stated in Section \ref{Conclusion}.

\section{Methodology\label{methodology}}

This study addresses the problem of an aerial payload transportation
system with human physical interaction using adaptive backstepping
control for position regulation, FNTSMC for attitude stabilization,
and an admittance controller for human interaction. The aerial system
lifts the transported payload while the human operator can interact
by applying force to the payload directly guiding the system in 3D
space. The interaction forces are measured using two force-torque
sensors attached at the point of contact between the quadrotors and
the payload. % We will introduce the dynamic model of the quadrotors and the payload separately; while the combined dynamic model for the entire assistive system will be used for control purposes. 

\subsection{System Model}

\begin{assum}\label{assum:Assum1} The following assumptions are
	considered: 
	\begin{itemize}
		\item The system comprises of three distinct rigid bodies: two identical
		quadrotors with known physical characteristics and a rigidly attached
		beam-shaped payload. 
		\item The system is symmetric along the $X$ and $Y$ axes. 
	\end{itemize}
\end{assum}
\vspace*{0.3cm}
For the modeling purposes, we assign a world inertial
reference frame $\mathcal{W}_{I}=\{O_{I},X_{I},Y_{I},Z_{I}\}$, body-attached
reference frames $L_{p}=\{o_{p},x_{p},y_{p},z_{p}\}$, and $q_{i}=\{o_{i},x_{q_{i}},y_{q_{i}},z_{q_{i}}\}$
attached to the CoM of the payload and the CoM each quadrotor, respectively,
where $(i=1,2)$. Let $\Omega=[p,q,r]^{\top}$ be the angular rates
defined in the body-attached frame which is the same for all components
of the system due to rigid connection in Assumptions \ref{assum:Assum1},
and $\Phi=[\phi,\theta,\psi]^{\top}$ be the Euler angles (roll, pitch,
and yaw) defined in $\mathcal{W}_{I}$ describing the system's orientation.
The rotation matrix from the body-attached frame to $\mathcal{W}_{I}$-frame
is $R\in SO(3)$, where $RR^{\top}=\mathbf{I}$ and $det(R)=+1$ with
$\mathbf{I}$ being the identity matrix \cite{hash2019_arXiv_Special_Survey,hash2021_AeScTe_SLAM}.
\begin{figure}
	\centerline{\includegraphics[scale=0.34]{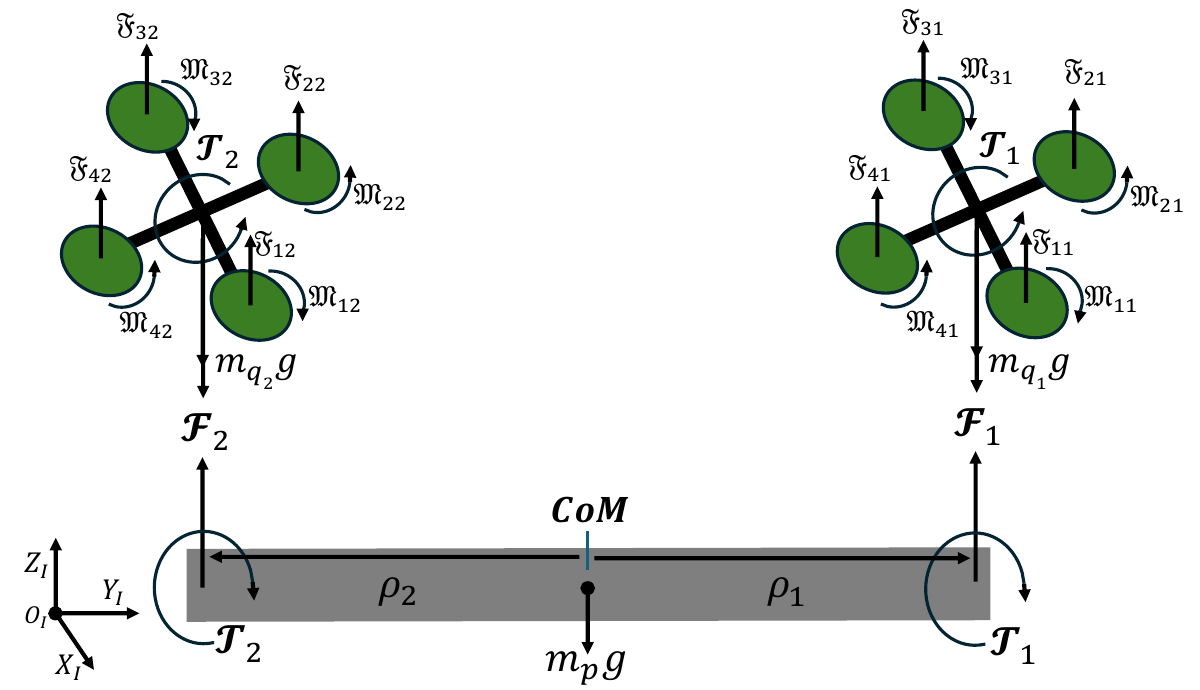}}
	\caption{Free body diagram of the entire system components.}
	\label{fig1} 
\end{figure}

\subsubsection{Quadrotor Dynamics}

Let $r_{q_{i}}=[x_{q_{i}},y_{q_{i}},z_{q_{i}}]^{\top}$ be the position
of the $i^{th}$ quadrotor and $\nu_{q_{i}}=\dot{r}_{q_{i}}$ defined
in $\mathcal{W}_{I}$. The underactuated UAV model dynamics are described
according to the Newton-Euler method as follows \cite{hashim2023exponentially,naser2025aerial}:
\begin{equation}
	\begin{aligned}m_{q_{i}}\dot{\nu}_{q_{i}} & =R\mathcal{F}_{qi}e_{z}-m_{q_{i}}ge_{z}-\mathcal{F}_{i}\\
		\mathcal{I}_{q_{i}}\dot{\Omega} & =\tau_{qi}-\Omega\times\mathcal{I}_{q_{i}}\Omega-\mathcal{T}_{i}
	\end{aligned}
	\label{trans_rot_quad}
\end{equation}
where $m_{q_{i}}\in\mathbb{R}$ and $\mathcal{I}_{q_{i}}\in\mathbb{R}^{3\times3}$
are the mass and moment of inertia, and $\mathcal{F}_{i}\in\mathbb{R}^{3}$
and $\mathcal{T}_{i}\in\mathbb{R}^{3}$ are the force and torque exerted
by the $i^{th}$ quadrotor on the attached payload, respectively,
$g\in\mathbb{R}$ denotes the gravity, $e_{z}=[0,0,1]^{\top}$ is
a unit vector along $z$ axis, and $\mathcal{F}_{qi}\in\mathbb{R}$
and $\tau_{qi}\in\mathbb{R}^{3}$ represents the thrust and the (rolling,
pitching, and yawing) moments measured in the body-attached frame,
such that: 
\begin{equation}
	\begin{bmatrix}\mathcal{F}_{qi}\\
		\tau_{qi}
	\end{bmatrix}=\begin{bmatrix}\mathfrak{F}_{i1}+\mathfrak{F}_{i2}+\mathfrak{F}_{i3}+\mathfrak{F}_{i4}\\
		d(\mathfrak{F}_{i2}-\mathfrak{F}_{i4})\\
		d(\mathfrak{F}_{i3}-\mathfrak{F}_{i1})\\
		\mathfrak{M}_{i1}-\mathfrak{M}_{i2}+\mathfrak{M}_{i3}-\mathfrak{M}_{i4}
	\end{bmatrix}\label{total_i_thrust_moments}
\end{equation}
where $\mathfrak{F}_{ij}=k_{f}\omega_{ij}^{2}$ and $\mathfrak{M}_{ij}=k_{m}\omega_{ij}^{2}$
are the thrust and moment of each rotor per quadrotor, and $k_{f},k_{m}$
are the thrust and moment constants, respectively, $\omega_{ij}$
is the rotor angular speed, and $d$ is the quadrotor's arm length.
\vspace*{0.3cm}

\subsubsection{Payload Dynamics}

Let $r_{p}=[x_{p},y_{p},z_{p}]^{\top}$ be the payload's CoM position
and $\nu_{p}=\dot{r}_{p}$ is defined in $\mathcal{W}_{I}$. Referring
to Fig. \ref{fig1}, the payload dynamics are formulated as follows:
\begin{equation}
	\begin{aligned} & m_{p}\dot{\nu}_{p}=(\mathcal{F}_{1}+\mathcal{F}_{2})-m_{p}ge_{z}\\
		& \mathcal{I}_{p}\dot{\Omega}~=(\mathcal{T}_{1}+\mathcal{T}_{2})-\Omega\times\mathcal{I}_{p}\Omega+(\rho_{1}\times\mathcal{F}_{1})+(\rho_{2}\times\mathcal{F}_{2})
	\end{aligned}
	\label{trans_rot_payload}
\end{equation}
where $\dot{\nu}_{p}\in\mathbb{R}^{3}$ is the translational acceleration,
$m_{p}\in\mathbb{R}$ and $\mathcal{I}_{p}\in\mathbb{R}^{3\times3}$
are the mass and moment of inertia, respectively. The terms $\rho_{1}\times\mathcal{F}_{1}$
and $\rho_{2}\times\mathcal{F}_{2}$ represent the moments exerted
by the $i^{th}$ quadrotor at the points of attachment, and $\rho_{i}=r_{q_{i}}-r_{p}$
represents the vectors between the payload's CoM and the $i^{th}$
quadrotor.
\vspace*{0.3cm}

\subsubsection{Entire Aerial Assistive System Dynamics}

Let $r_{s}=[x,y,z]^{\top}$ be the entire aerial system's CoM position
and $\nu_{s}=\dot{r}_{s}=[\nu_{s_{x}},\nu_{s_{y}},\nu_{s_{z}}]^{\top}$
is the velocity defined in $\mathcal{W}_{I}$. Based on Assumption
\ref{assum:Assum1} and Fig. \ref{fig1}, integrating the dynamics
of the quadrotors in \eqref{trans_rot_quad} with the payload dynamics
in \eqref{trans_rot_payload} and including the aerodynamic drag effects,
the dynamics of the entire system becomes: 
\begin{equation}
	\begin{aligned}m_{s}\dot{\nu}_{s} & =R\mathcal{U}_{th}e_{z}-m_{s}ge_{z}-F_{d}\in\mathbb{R}^{3}\\
		\mathcal{I}_{s}\dot{\Omega} & =\mathcal{U}_{m}-\Omega\times\mathcal{I}_{s}\Omega-M_{d}\in\mathbb{R}^{3}
	\end{aligned}
	\label{trans_rot_sys}
\end{equation}
where $\dot{\nu}_{s}\in\mathbb{R}^{3}$ is the translational acceleration,
$m_{s}=m_{q_{1}}+m_{q_{2}}+m_{p}\in\mathbb{R}$ and $\mathcal{I}_{s}\in\mathbb{R}^{3\times3}$
are the system total mass and moment of inertia, respectively, $F_{d}=K_{dl}\nu_{s}$
and $M_{d}=K_{dr}\Omega$ are the aerodynamic drag forces and torques,
$K_{dl},K_{dr}\in\mathbb{R}^{3\times3}$ are positive diagonal matrices
of the translational and rotational drag coefficients, respectively,
$\mathcal{U}_{th}\in\mathbb{R}$ and $\mathcal{U}_{m}\in\mathbb{R}^{3}$
are the total thrust and moments generated by the two quadrotors in
the body-attached frames. Since the generated thrust and moment of
each quadrotor measured in its own body-attached frame, these thrusts
and moments needs to be mapped to total thrust and moment of the entire
aerial system based on its configuration, such that: 
\begin{equation}
	\begin{bmatrix}\mathcal{U}_{th}\\
		\mathcal{U}_{m}
	\end{bmatrix}=\Lambda u_{d}\label{contro_input}
\end{equation}
where $u_{d}\in\mathbb{R}^{8}$ represents the control input's vector
for two quadrotors in the system as given below: 
\begin{equation}
	u_{d}=[\mathcal{F}_{q_{1}},\tau_{q_{11}},\tau_{q_{12}},\tau_{q_{13}},\mathcal{F}_{q_{2}},\tau_{q_{21}},\tau_{q_{22}},\tau_{q_{23}}]^{\top}\label{u_inputs}
\end{equation}
and $\Lambda\in\mathbb{R}^{4\times8}$ is a matrix that is constructed
based on the configuration of the aerial system, such that: 
\begin{equation}
	\Lambda=\begin{bmatrix}1 & 0 & 0 & 0 & 1 & 0 & 0 & 0\\
		\rho_{1}(2) & 1 & 0 & 0 & \rho_{2}(2) & 1 & 0 & 0\\
		-\rho_{1}(1) & 0 & 1 & 0 & -\rho_{2}(1) & 0 & 1 & 0\\
		0 & 0 & 0 & 1 & 0 & 0 & 0 & 1
	\end{bmatrix}\label{b_matrix}
\end{equation}
\vspace*{0.3cm}

\begin{rem}
	\label{rem_1} Firstly the total controller $[\mathcal{U}_{th},\mathcal{U}_{m}^{\top}]^{\top}$
	is designed and it is then mapped to find the control inputs of each
	quadrotor $u_{d}$ to track the desired human guidance. The system
	in \eqref{contro_input} is underdetermined which means that there
	are eight unknowns while there are only four equations. Hence, we
	need to optimize our solution through minimizing of a cost function
	$\mathcal{J}(u_{d}):\mathbb{R}^{8}\rightarrow\mathbb{R}$, such that: 
\end{rem}
\begin{equation}
	\mathcal{J}=\sum_{i=1}^{2}\sigma_{i1}\mathcal{F}_{q_{i}}+\sigma_{i2}\tau_{q_{i1}}+\sigma_{i3}\tau_{q_{i2}}+\sigma_{i4}\tau_{q_{i3}}\label{cost_fun}
\end{equation}
where $\sigma_{ij}$ are coefficients of $\mathcal{J}$ which are
used to find a matrix $\Gamma\in\mathbb{R}^{8\times8}$ such that
$\mathcal{J}$ in \eqref{cost_fun} can be rewritten as $\mathcal{J}=\|\Gamma u_{d}\|_{2}^{2}$,
and $\Gamma$ is defined as: 
\begin{equation}
	\Gamma=\sqrt{diag(\sigma_{11},\sigma_{12},\sigma_{13},\sigma_{14},\sigma_{21},\sigma_{22},\sigma_{23},\sigma_{24})}\label{h_matrix}
\end{equation}
Based on \eqref{b_matrix} and \eqref{h_matrix}, the optimized solution
is determined using pseudo inverse as \cite{babaie2017robust}: 
\begin{equation}
	\begin{split}u_{d}^{*} & =\text{argmin}\{\mathcal{J}|[\mathcal{U}_{th},~\mathcal{U}_{m}^{\top}]^{\top}=\Lambda u_{d}\}\\
		& =\Gamma^{-1}(\Lambda\Gamma^{-1})^{+}[\mathcal{U}_{th},~\mathcal{U}_{m}^{\top}]^{\top}\\
		& =\Gamma^{-2}\Lambda^{\top}(\Lambda\Gamma^{-2}\Lambda^{\top})^{-1}[\mathcal{U}_{th},~\mathcal{U}_{m}^{\top}]^{\top}
	\end{split}
	\label{optim_cont_final}
\end{equation}
where $+$ represents the pseudo inverse and the optimized solution
$u_{d}^{*}$ in \eqref{optim_cont_final} satisfies Remark \ref{rem_1}.
\begin{figure}[!hbt]
	\centering \includegraphics[scale=0.28]{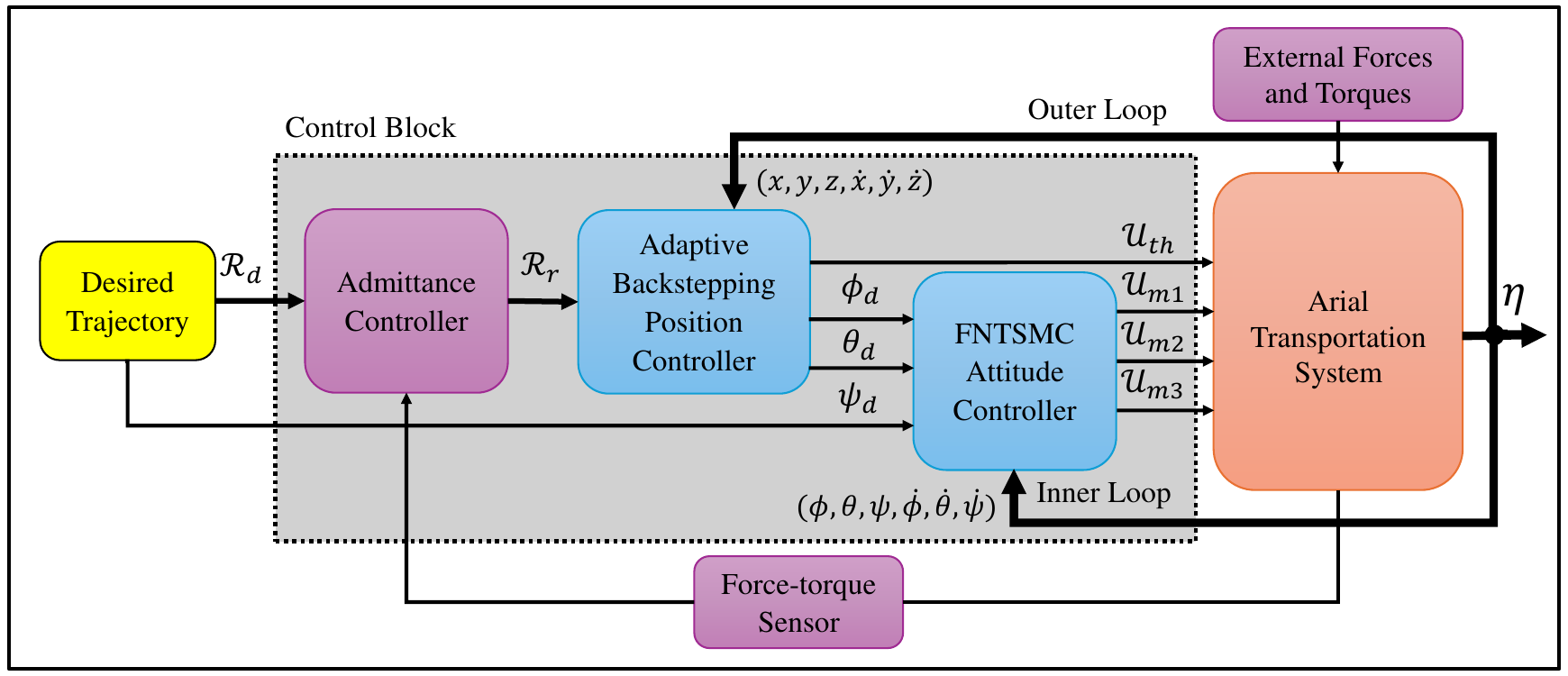} \caption{Illustrative diagram of the proposed control system.}
	\label{fig2} 
\end{figure}

\section{Controller Design and Stability Analysis \label{control_section}}

%This study proposes a hybrid control strategy combining Backstepping control for position regulation, FNTSMC for attitude stabilization, and an admittance controller for human interaction. The Backstepping method systematically designs controllers for nonlinear systems, while FNTSMC ensures robustness and rapid convergence. The admittance controller translates human-applied forces into reference trajectories, allowing precise guidance. This integrated approach enhances the performance and stability of the assistive aerial cooperative payload transportation system. 
Fig. \ref{fig2} shows a schematic diagram of the proposed control
system. The model dynamics in \eqref{trans_rot_sys} can be written
in state-space form for the purpose of control design as follows:
\begin{equation}
	\dot{\eta}=\mathcal{Q}(\eta)+b(\eta)\mathcal{U}\label{xsi}
\end{equation}

where $\mathcal{Q}(\eta)$ and $b(\eta)\neq0$ are nonlinear smooth
functions, $\mathcal{U}=[\mathcal{U}_{th},~\mathcal{U}_{m}]^{\top}\in\mathbb{R}^{4}$
is the control input, and $\eta=[x,\dot{x},y,\dot{y},z,\dot{z},\phi,\dot{\phi},\theta,\dot{\theta},\psi,\dot{\psi}]^{\top}\in\mathbb{R}^{12}$
is the state vector of the system. Assuming that for small angles
applications $[\dot{\phi},\dot{\theta},\dot{\psi}]^{\top}=[p,q,r]^{\top}$,
the dynamic equations in \eqref{xsi} is extended as follows: 
\begin{equation}
	\begin{aligned} & \dot{x}~~=~~\nu_{s_{x}}\\
		& \dot{\nu}_{s_{x}}=\left[-k_{dl}\nu_{s_{x}}+(s\theta c\psi+s\phi c\theta s\psi)\mathcal{U}_{th}\right]/m_{s}\\
		& \dot{y}~~=~~\nu_{s_{y}}\\
		& \dot{\nu}_{s_{y}}=\left[-k_{dl}\nu_{s_{y}}+(s\theta s\psi-s\phi c\theta c\psi)\mathcal{U}_{th}\right]/m_{s}\\
		& \dot{z}~~=~~\nu_{s_{z}}\\
		& \dot{\nu}_{s_{z}}=\left[-k_{dl}\nu_{s_{z}}-m_{s}g+(c\phi c\theta)\mathcal{U}_{th}\right]/m_{s}\\
		& \dot{\phi}~~=~~p\\
		& \ddot{\phi}~~=[-k_{dr}\dot{\phi}+(\mathcal{I}_{sy}-\mathcal{I}_{sz})\dot{\theta}\dot{\psi}+\mathcal{U}_{m1}]/\mathcal{I}_{sx}\\
		& \dot{\theta}~~=~~q\\
		& \ddot{\theta}~~=[-k_{dr}\dot{\theta}+(\mathcal{I}_{sz}-\mathcal{I}_{sx})\dot{\phi}\dot{\psi}+\mathcal{U}_{m2}]/\mathcal{I}_{sy}\\
		& \dot{\psi}~~=~~r\\
		& \ddot{\psi}~~=[-k_{dr}\dot{\psi}+(\mathcal{I}_{sx}-\mathcal{I}_{sy})\dot{\phi}\dot{\theta}+\mathcal{U}_{m3}]/\mathcal{I}_{sz}
	\end{aligned}
	\label{state_space_equ}
\end{equation}

\begin{rem}
	\label{rem:Remark1}From \eqref{state_space_equ} it can be realized
	that there are two coupled subsystems. The first one $(\dot{x}$ to
	$\dot{\nu}_{s_{z}})$ is the translation subsystem with one control
	input $(\mathcal{U}_{th})$ and three outputs $(x,y,z)$. The second
	subsystem $(\dot{\phi}$ to $\ddot{\psi})$ has three control inputs
	$(\mathcal{U}_{m1},\mathcal{U}_{m2},\mathcal{U}_{m3})$ and three
	outputs $(\phi,\theta,\psi)$. Therefore, we need to account for such
	problem in our controller design. 
\end{rem}

\subsection{Position Control using Adaptive Backstepping}

The position control is achieved via adaptive backstepping control
design which involves constructing Lyapunov functions to ensure stability
and designing virtual control laws through recursive steps to address
underactuation issue. Let $\eta_{d}=[x_{d},\dot{x}_{d},y_{d},\dot{y}_{d},z_{d},\dot{z}_{d},\phi_{d},\dot{\phi}_{d},\theta_{d},\dot{\theta}_{d},\psi_{d},\dot{\psi}_{d}]^{\top}\in\mathbb{R}^{12}$
be the desired reference trajectories of the system, generated by
the admittance controller according to the applied forces of the human
guidance. Let the position tracking error be: 
\begin{equation}
	E_{p}=\begin{bmatrix}E_{px}\\
		E_{py}\\
		E_{pz}
	\end{bmatrix}=\begin{bmatrix}x_{d}-x\\
		y_{d}-y\\
		z_{d}-z
	\end{bmatrix}\in\mathbb{R}^{3}\label{pos_error}
\end{equation}
The time derivative of \eqref{pos_error} is: 
\begin{equation}
	\dot{E}_{p}=\begin{bmatrix}\dot{E}_{px}\\
		\dot{E}_{py}\\
		\dot{E}_{pz}
	\end{bmatrix}=\begin{bmatrix}\dot{x}_{d}-\dot{x}\\
		\dot{y}_{d}-\dot{y}\\
		\dot{z}_{d}-\dot{z}
	\end{bmatrix}=\begin{bmatrix}\dot{x}_{d}-\nu_{s_{x}}\\
		\dot{y}_{d}-\nu_{s_{y}}\\
		\dot{z}_{d}-\nu_{s_{z}}
	\end{bmatrix}\label{pos_error_dot}
\end{equation}
Let $\odot$ and $\Pexp$ be the elementwise multiplication and exponentiation
operators, respectively. Based on \eqref{pos_error} and \eqref{pos_error_dot},
let the following vector of scalar functions, $V_{p}=\frac{1}{2}E_{p}^{\Pexp2}$
be Lyapunov function candidates where $V_{p}=[V_{px},V_{py},V_{pz}]^{\top}$.
Taking the time derivative of $V_{p}$, such that: 
\begin{equation}
	\dot{V}_{p}=E_{p}\odot\dot{E}_{p}=E_{p}\odot\begin{bmatrix}\dot{x}_{d}-\nu_{s_{x}}\\
		\dot{y}_{d}-\nu_{s_{y}}\\
		\dot{z}_{d}-\nu_{s_{z}}
	\end{bmatrix}\label{pos_lyapunov_dot}
\end{equation}
Consider $X_{temp}=[\nu_{s_{xd}},\nu_{s_{yd}},\nu_{s_{zd}}]^{\top}\in\mathbb{R}^{3}$
as a temporary desired control input used to ensure asymptotic stability
of the position, such that: 
\begin{equation}
	X_{temp}=K_{p}E_{p}+\dot{X}_{d}\in\mathbb{R}^{3}\label{temp_control}
\end{equation}
where $X_{d}=[x_{d},y_{d},z_{d}]^{\top}$ and $K_{p}\in\mathbb{R}^{3\times3}=diag(k_{p1},k_{p2},k_{p3})>0$.
Substituting \eqref{temp_control} into \eqref{pos_lyapunov_dot},
one gets: 
\begin{equation}
	\begin{aligned}\dot{V}_{p} & =E_{p}\odot(\dot{X}_{d}-X_{temp})=E_{p}\odot(\dot{X}_{d}-K_{p}E_{p}-\dot{X}_{d})\\
		& =-K_{p}E_{p}^{\Pexp2}<0,~~\forall\text{~elements of}~E_{p}\neq0.
	\end{aligned}
	\label{V_p_dot_result}
\end{equation}
Based on \eqref{V_p_dot_result}, the origin $(E_{p}=0)$ is globally
asymptotically stable. As a second step, let $E_{v}$ be the error
between the $X_{temp}$ and $X_{state}=[\nu_{s_{x}},\nu_{s_{y}},\nu_{s_{z}}]^{\top}\in\mathbb{R}^{3}$
from \eqref{state_space_equ} as follows: 
\begin{equation}
	E_{v}=\begin{bmatrix}E_{vx}\\
		E_{vy}\\
		E_{vz}
	\end{bmatrix}=(K_{p}E_{p}+\dot{X}_{d})-X_{state}\label{vel_error}
\end{equation}
Define $V_{v}=\frac{1}{2}E_{p}^{\Pexp2}+\frac{1}{2}E_{v}^{\Pexp2}$
as a vector of Lyapunov function candidates, where $V_{v}=[V_{vx},V_{vy},V_{vz}]^{\top}$;
take its time derivative, and substitute \eqref{vel_error} into the
result, such that: 
\begin{multline}
	\dot{V}_{v}=E_{p}\odot\dot{E}_{p}+E_{v}\odot\dot{E}_{v}=E_{p}\odot(\dot{X}_{d}-X_{temp})~~+\\
	E_{v}\odot[(K_{p}\dot{E}_{p}+\ddot{X}_{d})-\dot{X}_{state}]\label{vel_lyapunov_fun_dot1}
\end{multline}
Since we require the state variables $X_{state}$ to track the temporary
desired control input $X_{temp}$, therefore, equation \eqref{vel_lyapunov_fun_dot1}
can be rearranged and simplified as follows: 
\begin{equation}
	\dot{V}_{v}=-K_{p}E_{p}^{\Pexp2}+E_{v}\odot[-E_{p}+K_{p}\dot{E}_{p}+\ddot{X}_{d}-U_{v}-\mathcal{Q}_{p}]\label{vel_lyapunov_fun_dot2}
\end{equation}
where $U_{v}$ is a virtual control of the position subsystem from
\eqref{state_space_equ}, and it is given as follows: 
\begin{equation}
	U_{v}=\begin{bmatrix}u_{vx}\\
		u_{vy}\\
		u_{vz}
	\end{bmatrix}=\frac{1}{m_{s}}\begin{bmatrix}(s\theta c\psi+s\phi c\theta s\psi)\mathcal{U}_{th}\\
		(s\theta s\psi-s\phi c\theta c\psi)\mathcal{U}_{th}\\
		-m_{s}g+(c\phi c\theta)\mathcal{U}_{th}
	\end{bmatrix}\label{virtual_cont_equ}
\end{equation}
and $\mathcal{Q}_{p}=\frac{1}{m_{s}}[-k_{dl}\nu_{s_{x}},-k_{dl}\nu_{s_{y}},-k_{dl}\nu_{s_{z}}]^{\top}$
represents the drag forces in \eqref{state_space_equ}. The virtual
controller in \eqref{vel_lyapunov_fun_dot2} needs to be designed
to ensure the asymptotic stability of the system and the errors' convergence
to the origin as follows: 
\begin{equation}
	U_{v}=\hat{K}_{v}E_{v}-E_{p}+K_{p}\dot{E}_{p}+\ddot{X}_{d}-\mathcal{Q}_{p}\label{virtual_cont_design_equ}
\end{equation}
where $\hat{K}_{v}\in\mathbb{R}^{3\times3}=diag(\hat{k}_{v1},\hat{k}_{v2},\hat{k}_{v3})$
represents the the adaptive controller's parameter to estimate the
constant parameter $K_{v}\in\mathbb{R}^{3\times3}=diag(k_{v1},k_{v2},k_{v3})>0$
with the following adaptation law: 
\begin{equation}
	\dot{\hat{K}}_{v}=\mathcal{B}E_{v}^{\Pexp2}\label{adaptation_law}
\end{equation}
where $\mathcal{B}\in\mathbb{R}^{3\times3}=diag(\beta_{1},\beta_{2},\beta_{3})>0$.
To ensure the stability of the position subsystem with the adaptation
law in \eqref{adaptation_law}, let $V_{pv}\in\mathbb{R}^{3}$ be
a Lyapunov function candidates as follows: 
\begin{equation}
	V_{pv}=\frac{1}{2}E_{p}^{\Pexp2}+\frac{1}{2}E_{v}^{\Pexp2}+\frac{1}{2}\mathcal{B}^{-1}\tilde{K}_{v}^{2}\label{lyapunov_fun_pv}
\end{equation}
where $(\tilde{K}_{v}=K_{v}-\hat{K}_{v})$. Take the time derivative
of \eqref{lyapunov_fun_pv}, one gets: 
\begin{equation}
	\begin{split}\dot{V}_{pv} & =-K_{p}E_{p}^{\Pexp2}-\hat{K}_{v}E_{v}^{\Pexp2}+\mathcal{B}^{-1}\tilde{K}_{v}\dot{\tilde{K}}_{v}\\
		& =-K_{p}E_{p}^{\Pexp2}-(K_{v}-\tilde{K}_{v})E_{v}^{\Pexp2}-\mathcal{B}^{-1}\tilde{K}_{v}\dot{\hat{K}}_{v}\\
		& =-K_{p}E_{p}^{\Pexp2}-K_{v}E_{v}^{\Pexp2}+\tilde{K}_{v}(E_{v}^{\Pexp2}-\mathcal{B}^{-1}\dot{\hat{K}}_{v})
	\end{split}
	\label{vel_lyapunov_fun_dot3_final}
\end{equation}
Substituting \eqref{adaptation_law} into \eqref{vel_lyapunov_fun_dot3_final},
results in: 
\begin{equation}
	\dot{V}_{pv}=-K_{p}E_{p}^{\Pexp2}-K_{v}E_{v}^{\Pexp2}<0,~\forall~E_{p}\neq0~\&~E_{v}\neq0\label{vel_lyapunov_fun_dot_result}
\end{equation}
by which the stability of the position is guaranteed according to
the Lyapunov theory. Based on \eqref{virtual_cont_design_equ}, the
virtual controllers of the position subsystem are given as follows:
\begin{equation}
	\begin{aligned}u_{vx} & =-E_{px}+\hat{k}_{v1}E_{vx}+k_{p1}\dot{E}_{px}+\ddot{x}_{d}-\mathcal{Q}_{p}(1)\\
		u_{vy} & =-E_{py}+\hat{k}_{v2}E_{vy}+k_{p2}\dot{E}_{py}+\ddot{y}_{d}-\mathcal{Q}_{p}(2)\\
		u_{vz} & =-E_{pz}+\hat{k}_{v3}E_{vz}+k_{p3}\dot{E}_{pz}+\ddot{z}_{d}-\mathcal{Q}_{p}(3)
	\end{aligned}
	\label{designed_virtual_cont_equ}
\end{equation}
Based on \eqref{virtual_cont_equ}, the total thrust of the entire
system $\mathcal{U}_{th}$ and the desired orientation angles $(\phi_{d},\theta_{d})$
as outputs of the (outer loop) position controller are given as follows:
\begin{equation}
	\begin{aligned} & \mathcal{U}_{th}=m_{s}\sqrt{u_{vx}^{2}+u_{vy}^{2}+(u_{vx}+g)^{2}}\\
		& \theta_{d}=\arctan\left(\frac{u_{vx}\cos(\psi_{d})+u_{vy}\sin(\psi_{d})}{u_{vz}+g}\right)\\
		& \phi_{d}=\arctan\left(\frac{\cos(\theta_{d})(u_{vx}\sin(\psi_{d})-u_{vy}\cos(\psi_{d}))}{u_{vz}+g}\right)
	\end{aligned}
	\label{equ_des_orient}
\end{equation}

\subsection{Attitude Subsystem Controller}

The attitude control of the system is managed by FNTSMC. This controller
offers rapid convergence and robustness against uncertainties. The
controller design process and the stability analysis according to
Lyapunov theory are given follows: 

\subsubsection{Sliding surface design}

Let $E_{\Phi}\in\mathbb{R}^{3}$ be the tracking error of the attitude
subsystem which is given as: 
\begin{equation}
	E_{\Phi}=\begin{bmatrix}E_{\phi}\\
		E_{\theta}\\
		E_{\psi}
	\end{bmatrix}=\eta_{d\Phi}-\eta_{\Phi}=\begin{bmatrix}\phi_{d}-\phi\\
		\theta_{d}-\theta\\
		\psi_{d}-\psi
	\end{bmatrix}\label{att_error}
\end{equation}
Take the time derivative of \eqref{att_error}, we get: 
\begin{equation}
	\dot{E}_{\Phi}=\dot{\eta}_{d\Phi}-\dot{\eta}_{\Phi}=\begin{bmatrix}\dot{\phi}_{d}-\dot{\phi}\\
		\dot{\theta}_{d}-\dot{\theta}\\
		\dot{\psi}_{d}-\dot{\psi}
	\end{bmatrix}\label{att_error_dot}
\end{equation}
For the system in \eqref{xsi} and the tracking error in \eqref{att_error},
the FNTSM surfaces of the attitude subsystem are chosen to be as follows:
\begin{equation}
	S_{\Phi}=\dot{E}_{\Phi}+\zeta E_{\Phi}+\gamma sgn(E_{\Phi})^{\varepsilon}\label{sliding_surface}
\end{equation}
where $S_{\Phi}=[S_{\phi},S_{\theta},S_{\psi}]^{\top}\in\mathbb{R}^{3}$,
$\zeta>0$, ~$\gamma>0$, ~$\varepsilon\geq1$, $sgn(E_{\Phi})^{\varepsilon}=|E_{\Phi}|^{\varepsilon}sgn(E_{\Phi})$
and its derivative is given as \cite{ref28} $\frac{d}{dt}(sgn(E_{\Phi})^{\varepsilon})=\varepsilon|E_{\Phi}|^{\varepsilon-1}\dot{E}_{\Phi}$,
and $sgn(\cdot)$ is the sign function. The sliding surface time derivative
in \eqref{sliding_surface} is: 
\begin{equation}
	\begin{split}\dot{S}_{\Phi} & =\ddot{E}_{\Phi}+\zeta\dot{E}_{\Phi}+\gamma\varepsilon|E_{\Phi}|^{\varepsilon-1}\dot{E}_{\Phi}\\
		& =\ddot{\eta}_{d\Phi}-\ddot{\eta}_{\Phi}+\zeta\dot{E}_{\Phi}+\gamma\varepsilon|E_{\Phi}|^{\varepsilon-1}\dot{E}_{\Phi}
	\end{split}
	\label{sliding_surface_dot}
\end{equation}
Substituting \eqref{xsi} into \eqref{sliding_surface_dot} results
in: 
\begin{equation}
	\dot{S}_{\Phi}=\ddot{\eta}_{d\Phi}-\mathcal{Q}_{\Phi}(\eta)-b(\eta)\mathcal{U}+\left(\zeta+\gamma\varepsilon|E_{\Phi}|^{\varepsilon-1}\right)\dot{E}_{\Phi}\label{sdot}
\end{equation}
Considering $\hat{\mathcal{U}}$ as an equivalent controller (continuous
control law) to achieve $\dot{S}_{\Phi}=0$ consider the following
expression: 
\begin{equation}
	\hat{\mathcal{U}}=\frac{1}{b(\eta)}\left[\ddot{\eta}_{d\Phi}-\mathcal{Q}_{\Phi}(\eta)+(\zeta+\gamma\varepsilon|E_{\Phi}|^{\varepsilon-1})\dot{E}_{\Phi}\right]\label{u_hat}
\end{equation}
To address the dynamic uncertainties and external disturbances, as
well as to increase the speed convergence to the sliding surface,
the control law can be formulated by incorporating a discontinuous
term into $\hat{\mathcal{U}}$ in \eqref{u_hat} as follows: 
\begin{equation}
	\mathcal{U}=\hat{\mathcal{U}}+\frac{1}{b(\eta)}\left[\kappa_{1}S_{\Phi}+\kappa_{2}sgn(S_{\Phi})\right]\label{control_final}
\end{equation}
where $\kappa_{1}>0$ and $\kappa_{2}>0$. The attitude controller
takes the desired roll and pitch angles $(\phi_{d}$ and $\theta_{d})$
that are generated by the position controller in \eqref{equ_des_orient}.
The desired yaw angle $\psi_{d}$ is set by the designer in view of
the entire system configuration to generate the control law in \eqref{control_final}
as follows: 
\begin{multline*}
	\mathcal{U}_{m1}=\mathcal{I}_{sx}\big[\ddot{\phi}_{d}-\mathcal{Q}_{\Phi}(1)+\left(\zeta_{\phi}+\gamma\varepsilon|E_{\phi}|^{\varepsilon-1}\right)\dot{E}_{\phi}~+\\
	\kappa_{1}S_{\phi}+\kappa_{2}sgn(S_{\phi})\big]
\end{multline*}
\begin{multline*}
	\mathcal{U}_{m2}=\mathcal{I}_{sy}\big[\ddot{\theta}_{d}-\mathcal{Q}_{\Phi}(2)+\left(\zeta_{\theta}+\gamma\varepsilon|E_{\theta}|^{\varepsilon-1}\right)\dot{E}_{\theta}~+\\
	\kappa_{1}S_{\theta}+\kappa_{2}sgn(S_{\theta})\big]
\end{multline*}
\begin{multline}
	\mathcal{U}_{m3}=\mathcal{I}_{sz}\big[\ddot{\psi}_{d}-\mathcal{Q}_{\Phi}(3)+\left(\zeta_{\psi}+\gamma\varepsilon|E_{\psi}|^{\varepsilon-1}\right)\dot{E}_{\psi}~+\\
	\kappa_{1}S_{\psi}+\kappa_{2}sgn(S_{\psi})\big]\label{orient_cont_equ}
\end{multline}
where 
\[
\mathcal{Q}_{\Phi}=\begin{bmatrix}\frac{1}{\mathcal{I}_{sx}}\left[-k_{dr}\dot{\phi}+(\mathcal{I}_{sy}-\mathcal{I}_{sz})\dot{\theta}\dot{\psi}\right]\\
	\frac{1}{\mathcal{I}_{sy}}\left[-k_{dr}\dot{\theta}+(\mathcal{I}_{sz}-\mathcal{I}_{sx})\dot{\phi}\dot{\psi}\right]\\
	\frac{1}{\mathcal{I}_{sz}}\left[-k_{dr}\dot{\psi}+(\mathcal{I}_{sx}-\mathcal{I}_{sy})\dot{\phi}\dot{\theta}\right]
\end{bmatrix}
\]

\subsection{Stability Analysis}
\begin{thm}
	\label{them1}Given the system dynamics outlined in \eqref{xsi} and
	the FNTSM surfaces specified in \eqref{sliding_surface}, the proposed
	control law in \eqref{control_final} guarantees asymptotic stability
	of the closed-loop system. Additionally, the system trajectories can
	reach to the equilibrium state on the FNTSM $(S_{\Phi}=0)$ within
	a finite time. 
\end{thm}
\textbf{Proof}: To prove Theorem \ref{them1}, let $V_{\Phi}$ be
a Lyapunov function candidate as follows: 
\begin{equation}
	V_{\Phi}=\frac{1}{2}S_{\Phi}^{\Pexp2},~~~~(V_{\Phi}\geqslant0~\forall\text{~elements of~}S_{\Phi})\label{lyap_fun}
\end{equation}
The time derivative of \eqref{lyap_fun} is: 
\begin{equation}
	\dot{V}_{\Phi}=S_{\Phi}\odot\dot{S}_{\Phi}\label{vdot}
\end{equation}
Substituting \eqref{sdot} into \eqref{vdot} results in the following:
\begin{equation}
	\dot{V}_{\Phi}=\left[\ddot{\eta}_{d\Phi}-\mathcal{Q}(\eta)-b(\eta)\mathcal{U}+\left(\zeta+\gamma\varepsilon|E_{\Phi}|^{\varepsilon-1}\right)\dot{E}_{\Phi}\right]\odot S_{\Phi}\label{vsdot}
\end{equation}
Substituting \eqref{control_final} into \eqref{vsdot} results in
the following: 
\begin{equation}
	\begin{split}\dot{V}_{\Phi} & =\left[-\kappa_{1}S_{\Phi}-\kappa_{2}sgn(S_{\Phi})\right]\odot S_{\Phi}\\
		& =-\kappa_{1}S_{\Phi}^{\Pexp2}-\kappa_{2}|S_{\Phi}|\leqslant0~,\forall S_{\Phi}
	\end{split}
	\label{vsdot_f}
\end{equation}
which ensures asymptotic convergence. To prove convergence to the
sliding surfaces $(S_{\Phi}=0)$ in a finite time, define $t_{r}$
as the reaching time and let us rewrite \eqref{vsdot_f} as follows:
\begin{equation}
	\dot{V}_{\Phi}=\frac{dV_{\Phi}}{dt}=-\kappa_{1}S_{\Phi}^{\Pexp2}-\kappa_{2}|S_{\Phi}|\label{vdot_tr}
\end{equation}
Using \eqref{lyap_fun}, The expression in \eqref{vdot_tr} can be
rewritten as follows: 
\begin{equation}
	\frac{dV_{\Phi}}{dt}\leqslant-2\kappa_{1}V_{\Phi}-\kappa_{2}\sqrt{2V_{\Phi}},~\text{s.t.}~dt\leqslant\frac{-dV_{\Phi}}{2\kappa_{1}V_{\Phi}+\alpha V_{\Phi}^{1/2}}\label{equ_dt}
\end{equation}
where $\alpha=\sqrt{2}\kappa_{2}$, then integrate both sides of \eqref{equ_dt}
such that: 
\begin{equation}
	\int_{0}^{t_{r}}dt\leqslant\int_{V_{\Phi}(0)}^{V_{\Phi}(t_{r})}\frac{-dV_{\Phi}}{2\kappa_{1}V_{\Phi}+\alpha V_{\Phi}^{1/2}}\label{equ_dt_int}
\end{equation}
Integrating \eqref{equ_dt_int} shows that the reaching time $(t_{r})$
can be calculated as: 
\begin{equation}
	t_{r}\leqslant\frac{1}{\kappa_{1}}\ln\left|\frac{2\kappa_{1}V_{\Phi}(0)^{1/2}+\alpha}{\alpha}\right|\label{reach_time}
\end{equation}
Based on \eqref{reach_time}, the system trajectories will reach $(S_{\Phi}=0)$
and the errors will converge to the origin within a finite time.

\subsection{Admittance Controller for Human Physical Interaction \label{controller_design}}

The admittance controller facilitates physical interaction between
the human operator and the aerial system. It measures the interaction
forces and generates reference trajectories for accurate tracking.
The admittance controller simulates a mass-damper-spring system by
modifying virtual inertia, damping, and stiffness of the robotic system
as given in the following mathematical relationship \eqref{equ4}:
\begin{equation}
	\mathcal{M}(\ddot{\mathcal{R}}_{d}-\ddot{\mathcal{R}}_{r})+\mathcal{C}(\dot{\mathcal{R}}_{d}-\dot{\mathcal{R}}_{r})+\mathcal{K}(\mathcal{R}_{d}-\mathcal{R}_{r})=\mathcal{F}_{h}\label{equ4}
\end{equation}
where: $\mathcal{M}\in\mathbb{R}^{3\times3},~\mathcal{C}\in\mathbb{R}^{3\times3},~\mathcal{K}\in\mathbb{R}^{3\times3}$
are diagonal matrices of the virtual mass, damping, stiffness, respectively,
$\mathcal{R}_{d}\in\mathbb{R}^{3}$ is the desired trajectory, $\mathcal{R}_{r}\in\mathbb{R}^{3}$
is the reference trajectory which will be generated by the admittance
controller according to the human operator's guidance through the
applied forces $\mathcal{F}_{h}\in\mathbb{R}^{3}$. %as shown in Fig. \ref{fig3}. 
%\begin{figure}[ht!]
%	\centering \includegraphics[width=0.65\linewidth]{external_force}
%	\caption{Interaction force and reference trajectories.}
%	\label{fig3} 
%\end{figure}
The goal for the aerial system is to hold its position when no external
force is applied. Consequently, $\ddot{\mathcal{R}}_{d}$ and $\dot{\mathcal{R}}_{d}$
are set to zero. In addition, the response of the physical interaction
can be adjusted by modifying the virtual stiffness $\mathcal{K}$.
A higher value of $\mathcal{K}$ results in a stiffer response, whereas
setting $\mathcal{K}$ to zero ensures full-compliance with the applied
external force. Furthermore, to mitigate environmental disturbances,
the admittance controller is activated when the measured force exceeds
a defined threshold.

\section{Simulation and Numerical Results\label{results-discussion}}

\begin{table}[t]
	\centering{}\caption{System and Controller parameters.}
	\begin{tabular}{llr}
		\hline 
		Symbol  & Definition  & Value/ Unit\tabularnewline
		\hline 
		\hline 
		$m_{q_{i}},m_{p},m_{s}$  & Masses (quad., load, total)  & $(1.4,0.45,3.25)/kg$\tabularnewline
		$d$  & Arm length of $i^{th}$ quad.  & $0.225/m$\tabularnewline
		$l_{p}$  & Length of payload  & $2.2/m$\tabularnewline
		$g$  & Gravitational acceleration  & $9.81/m/s^{2}$\tabularnewline
		$\mathcal{I}_{sx}$  & Moment of inertia in x-axis  & $3.039/kg~m^{2}$\tabularnewline
		$\mathcal{I}_{sy}$  & Moment of inertia in y-axis  & $0.051/kg~m^{2}$\tabularnewline
		$\mathcal{I}_{sz}$  & Moment of inertia in z-axis  & $3.072/kg~m^{2}$\tabularnewline
		$k_{dl}$  & Drag coef. along x,y,z-axes  & $55\times10^{-4}/N~s/m$\tabularnewline
		$k_{dr}$  & Drag coef. about x,y,z-axis  & $55\times10^{-4}/N~s$\tabularnewline
		$k_{p1},k_{p2},k_{p3}$  & Control constants  & 18, 9, 18\tabularnewline
		$\beta_{1},\beta_{2},\beta_{3}$  & Control constants  & 0.4, 0.4, 0.4\tabularnewline
		$\zeta_{\phi},\zeta_{\theta},\zeta_{\psi}$  & Control constants  & 22, 30, 22\tabularnewline
		$\varepsilon,\gamma,\kappa_{1},\kappa_{2}$  & Control constants  & 2, 5, 85, 55\tabularnewline
		$\mathcal{M}$  & Virtual mass  & $0.95/kg$\tabularnewline
		$\mathcal{C}$  & Virtual damping coef.  & $1.54/N~s/m$\tabularnewline
		$\mathcal{K}$  & Virtual spring constant  & $0/N~s/m$\tabularnewline
		\hline 
	\end{tabular}\label{tab:entire_sys_parameters} 
\end{table}

\begin{figure}[hbt!]
	\begin{centering}
		\subfloat[]{\begin{centering}
				\includegraphics[scale=0.5]{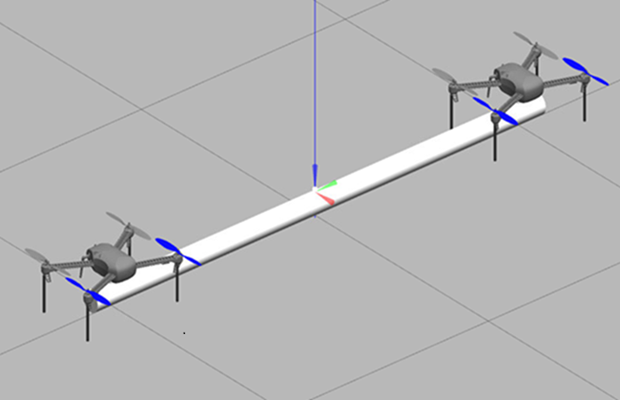}
				\par\end{centering}
		} \subfloat[]{\begin{centering}
				\includegraphics[scale=0.5]{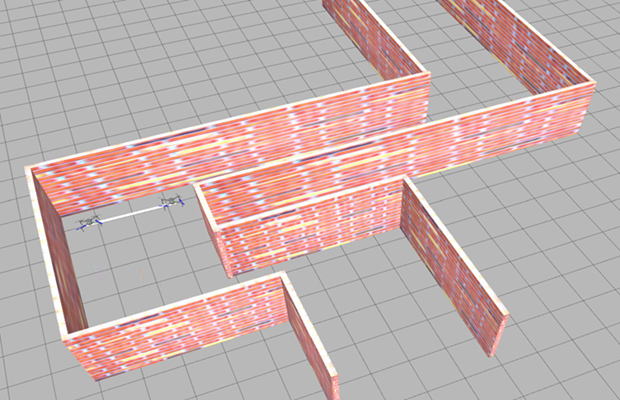}
				\par\end{centering}
		} 
		\par\end{centering}
	\centering{}\subfloat[]{\begin{centering}
			\includegraphics[width=0.9\linewidth]{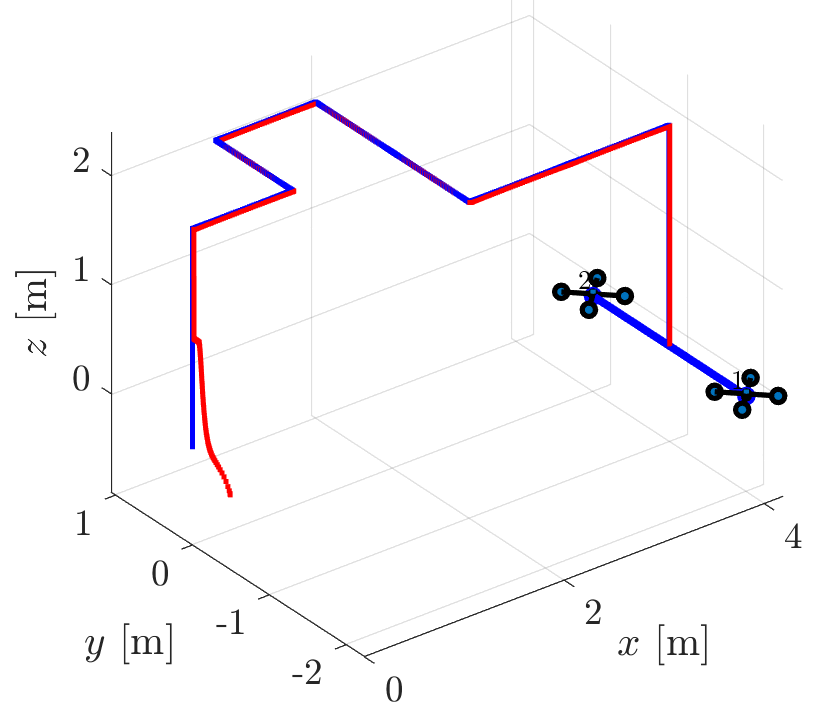}
			\par\end{centering}
	} \caption{Simulation; (a) The aerial system built in ROS and Gazebo 11, (b)
		Zigzag corridor as a navigation workplace, (c) MATLAB simulation with
		tracking path in 3D space.}
	\label{fig4} 
\end{figure}

The control design and the performance of the system which consists
of two quadrotors attached to a common payload as shown in Fig. \ref{fig4}.(a)
are validated using MATLAB (2024b), Robotic Operation System (ROS)
noetic, and Gazebo 11 (see Fig. \ref{fig4}). %The goal is to evaluate effectiveness of the proposed controllers in Section \ref{control_section} in stabilizing the system while tracking the desired human guidance.
Table \ref{tab:entire_sys_parameters} presents the parameters of
the system and the controller. Initially, the aerial vehicle lifted
the payload to the operating height, at which the admittance controller
was triggered to be ready for physical interaction. The user applied
an upward force to properly position the aerial vehicle. Subsequently,
simulated human guidance was performed by applying forces in various
directions to transport the payload to the final destination. The
results indicate that the system was fully complied with the applied
human guidance while the controller maintained stability throughout
the process until the final destination was reached, as shown in Fig.
\ref{fig4}.(b) and \ref{fig4}.(c). 
\begin{figure*}[!ht]
	\centering \includegraphics[width=0.85\linewidth]{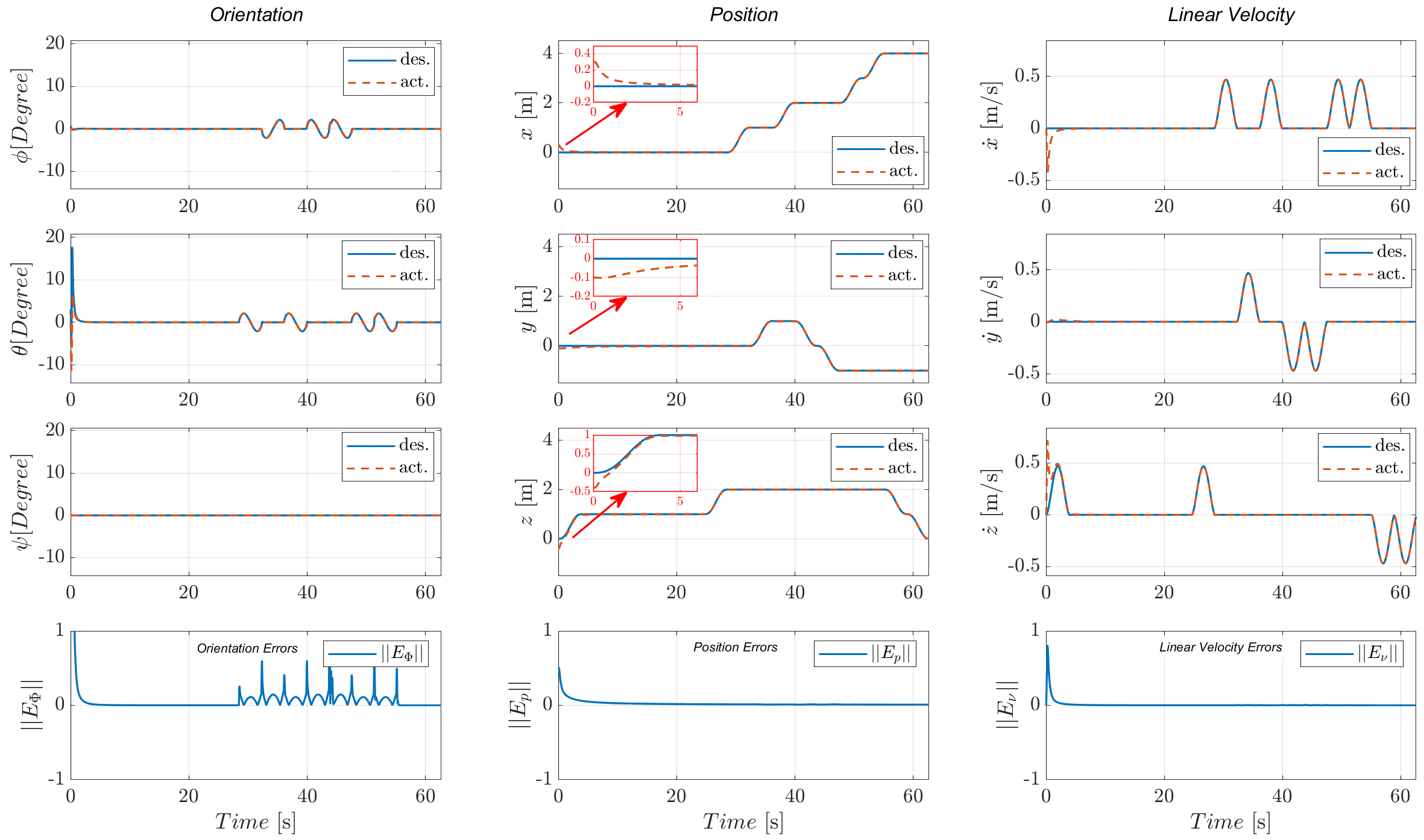}
	\caption{The simulation results; First column for the desired and actual orientation,
		Second column for the desired and actual position, Third column for
		the linear velocity in $(x,y,z)$ directions, and the bottom row for
		the normalized tracking errors.}
	\label{fig5} 
\end{figure*}

\begin{figure}[htbp]
	\centering \includegraphics[width=0.95\linewidth]{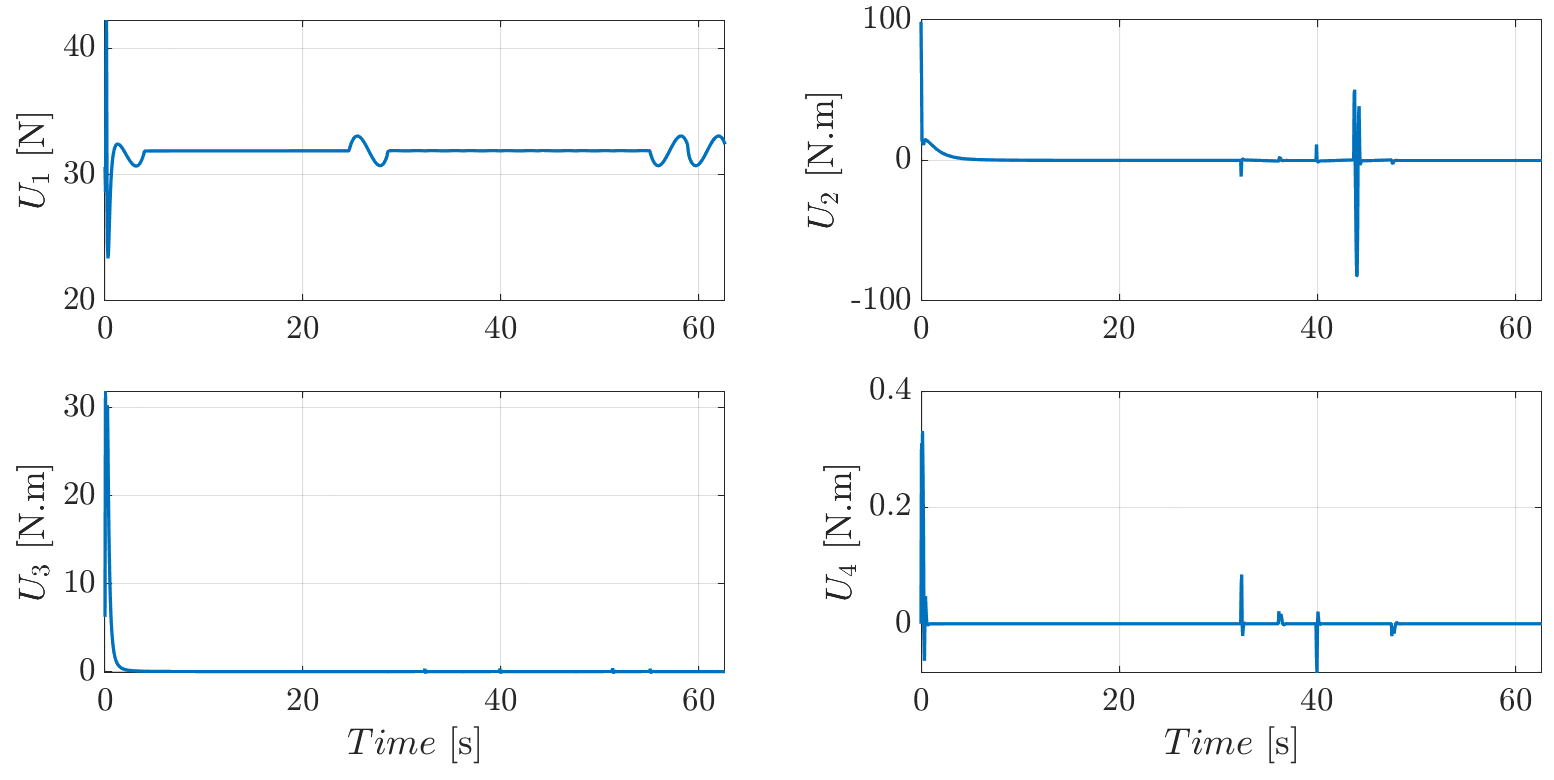}
	\caption{The control inputs.}
	\label{fig6} 
\end{figure}

Fig. \ref{fig5} shows the desired and actual trajectories of the
orientation, position, velocity, and tracking errors of the system.
The system successfully tracked the reference trajectories that generated
by the admittance controller as inputs to the position and attitude
controllers. The first column of Fig. \ref{fig5} shows the desired
and actual orientation of the system. It can be clearly seen that
the position controller managed to generate the desired orientation
to the attitude controller while the latter tracked this desired orientation
required to move the system to the positions in the $x$ and $(y)$
directions according to human guidance. In addition, The $(z)$ direction
graph illustrates that the system's altitude increased to the desired
altitude, maintained that altitude until the final destination, and
then the aerial system was landed. Meanwhile, the velocity graph displays
the corresponding required velocities. Fig. \ref{fig5} further illustrates
the normalized tracking errors for the system's position and attitude
controllers at the last row of the graph. The results indicate that
the position and velocity errors were reduced to zero throughout the
simulation, demonstrating the robustness of the proposed controller.
In contrast, the orientation error exhibited some fluctuations, likely
due to the high rolling and pitching control inputs at the start and
stop of the system, which are related to the system's high inertia.
However, the controller successfully bounded and reduced the orientation
error to zero. Fig. \ref{fig6} displays the control inputs generated
by the controller. The graph of the total thrust $\mathcal{U}_{th}$
shows the required generated thrust to reach the desired altitude
and maintaining the same thrust for the desired altitude, then finally
generated reverse thrust for landing. The other three graphs show
the rolling, pitching, and yawing moments which are required to navigate
to the desired positions in 3D space according to human guidance.

\section{Conclusion\label{Conclusion}}

This paper presents control strategy for an aerial cooperative payload
transportation system using two quadrotor UAVs rigidly connected to
a payload with human physical interaction. The integration of adaptive
backstepping control for the position subsystem, Fast Nonsingular
Terminal Sliding Mode Control (FNTSMC) for the attitude subsystem,
and an admittance controller for human interaction ensures precise
trajectory tracking. The proposed control design ensures position
and attitude asymptotic stabilization with responsive human guidance.
Simulation results demonstrate the effectiveness of the proposed method
in maintaining system stability and robust performance. % Future work will focus on experimental validation, estimation of the interaction forces, and adding adaptivity for unknown masses of different payloads.

\balance
\bibliographystyle{IEEEtran}
\bibliography{references}

\end{document}